# PanopMamba: Vision State Space Modeling for Nuclei Panoptic Segmentation

Ming Kang , *Member, IEEE*, Fung Fung Ting , Raphaël C.-W. Phan , *Senior Member, IEEE*, Zongyuan Ge , *Senior Member, IEEE*, Chee-Ming Ting , *Senior Member, IEEE*

*Abstract* — Nuclei panoptic segmentation supports cancer diagnostics by integrating both semantic and instance segmentation of different cell types to analyze overall tissue structure and individual nuclei in histopathology images. Major challenges include detecting small objects, handling ambiguous boundaries, and addressing class imbalance. To address these issues, we propose Panop-Mamba, a novel hybrid encoder-decoder architecture that integrates Mamba and Transformer with additional feature-enhanced fusion via state space modeling. We design a multiscale Mamba backbone and a State Space Model (SSM)-based fusion network to enable efficient long-range perception in pyramid features, thereby extending the pure encoder-decoder framework while facilitating information sharing across multiscale features of nuclei. The proposed SSM-based feature-enhanced fusion integrates pyramid feature networks and dynamic feature enhancement across different spatial scales, enhancing the feature representation of densely overlapping nuclei in both semantic and spatial dimensions. To the best of our knowledge, this is the first Mamba-based approach for panoptic segmentation. Additionally, we introduce alternative evaluation metrics, including image-level Panoptic Quality (iPQ), boundary-weighted PQ (wPQ), and frequency-weighted PQ (fwPQ), which are specifically designed to address the unique challenges of nuclei segmentation and thereby mitigate the potential bias inherent in vanilla PQ. Experimental evaluations on two multiclass nuclei segmentation benchmark datasets, MoNuSAC2020 and NuInsSeg, demonstrate the superiority of PanopMamba for nuclei panoptic segmentation over state-of-the-art methods. Ablation studies further validate the effectiveness of each component in improving performance. Consequently, the robustness of PanopMamba is validated across various metrics, while the distinctiveness of PQ variants is also demonstrated. Code is available at https://github.com/mkang315/PanopMamba.

*Index Terms*—Microscopy image analysis, computa-

tional pathology, histological image segmentation, nucleus segmentation, state space models

## I. INTRODUCTION

NUCLEI segmentation is a critical task in histopathology image analysis, particularly for analyzing tumor cell immune microenvironments and interpretable features of diagnostic and prognostic cancer indicators. It involves identifying and delineating critical cell nuclei in a tissue image, often employing Hematoxylin and Eosin (H&E) staining to enhance contrast. Nuclei panoptic segmentation offers valuable insights into understanding both the global structure and individual nuclei by combining semantic and instance segmentation for various types of cell nuclei within a single image. Nuclei panoptic segmentation remains challenging due to overlapping and densely packed nuclei in images, as well as the substantial variations in size, shape, and staining intensity. First, since the nuclei remain small even at high magnifications, the commonly used Intersection over Union (IoU) is a highly sensitive metric, often yielding low scores, even when the segmentations fall well within the expected variability in expert annotations [1]. Second, predicting ambiguous nuclear boundaries is more challenging than segmenting the internal area, as inconsistent boundary delineation makes performance evaluation around the object boundaries difficult. In contrast, IoU uniformly assesses all pixels in the image, as noted by [2]. Third, class/size imbalance in some histopathology datasets can bias performance metrics, as underrepresented nuclei categories contribute disproportionately to overall scores.

The current architecture of panoptic segmentation models in the generic domain is primarily based on a Transformer-based image encoder and a mask decoder [3], which differs from traditional CNN-based models that use a pure CNN backbone and segmentation heads. Panoptic SegFormer [4] and Mask2Former [5] first integrated the Transformer encoder with the Transformer decoder, incorporating masked attention, and achieved strong performance across major panoptic segmentation benchmarks on natural images, thanks to their enhanced feature extraction and mask prediction capabilities. MP-Former [6] employed multilayer mask-piloted training for Mask2former to overcome inconsistent mask predictions between consecutive decoder layers, and reduced failure cases

M. Kang, F. F. Ting, R. C.-W. Phan, and C.-M. Ting are with the School of Information Technology, Monash University, Malaysia Campus, Subang Jaya 47500, Malaysia (e-mails: ming.kang@monash.edu, ting.fung@monash.edu, raphael.phan@monash.edu, ting.cheeming@monash.edu).

Z. Ge and M. Kang are with the Augmented Intelligence and Multimodal analytics (AIM) for Health Lab, Faculty of Information Technology, Monash University, Clayton VIC 3800, Australia, and also with the Monash Medical AI Group, Monash University, Clayton VIC 3800, Australia (e-mail: zongyuan.ge@monash.edu).

Digital Object Identifier:

by using noised ground truth masks. OneFormer [7] extends the Mask2former with unified task-conditioned query representations within a single framework. OMG-Seg [8] adopted a frozen backbone and a shared multitask decoder, similar to Mask2former. In the histopathology domain, only one or two Convolutional Neural Network (CNN)-based models [9], [10] have been developed for nuclei panoptic segmentation. MaskConver [11] is a pure convolution model and consists of a CNN backbone, a pixel decoder with ConvNeXt-UNet, and a mask embedding generator that is similar to mask embedding classification in MaskFormer [12].

VMamba [13] extends the State Space Model (SSM) module in Mamba from one-dimensional (1D) sequences to two-dimensional (2D) inputs, such as images and videos, and inherits the advantages of capturing long-range dependencies with memory efficiency. VMamba employs a visual state space block with a 2D selective scan module, replacing the SSM+Selection module in the hierarchical network framework. In this framework, 2D inputs are first sliced into patches and then flattened into a patch sequence, without incorporating additional positional embeddings. Subsequently, Multi-Scale VMamba (MSVMamba) [14] replaced the 2D selective scan module in VMamba with a Multi-Scale 2D Scanning (MS2D) strategy to overcome the long-range forgetting issue of SSMs with limited parameters in computer vision. A recent work [15] introduced PathMamba for nuclei semantic segmentation for histopathology images. It is an extension of CNN-based methods that employ a Mamba [16], [17] decoder. However, the applications of vision state-space modeling to panoptic segmentation tasks remain largely unexplored in both the natural and medical imaging domains.

In this paper, we propose a novel hybrid architecture called PanopMamba for panoptic segmentation, which integrates a Mamba-based encoder with a Transformer decoder. We further incorporate a feature fusion network with state space modeling within the encoder-decoder framework to facilitate multi-scale fusion. To the best of our knowledge, this is the first work on Mamba/SSM-based panoptic segmentation for histopathology images and, more generally, natural images. The contributions of this work are summarized as follows:

1) We propose a novel hybrid neural network integrating Mamba and Transformer in an encoder-decoder architecture, which leverages the capability of Mamba for vision representations and the Transformer for mask attention. The Mamba-based encoder captures dependencies efficiently without the quadratic complexity of the Transformer. The Transformer pixel decoder with mask attention refines features across multiple scales to improve object detection and segmentation.

2) We design an SSM-based feature fusion network in the encoder-decoder architecture, which is a feature pyramid enhanced network that incorporates state space modeling from Mamba and CNNs with attention mechanisms to facilitate multiscale feature fusion. This allows aggregation of long-range information in the global features captured by the state space modeling, with the local features extracted by CNNs. This improves the segmentation of small objects of varying sizes and shapes.

3) We introduce alternative metrics to evaluate panoptic segmentation tailored to small-sized objects, ambiguous nuclei boundaries, and class imbalance of nuclei instances. Evaluation of two publicly available nuclear segmentation benchmark datasets reveals the superiority of the proposed PanopMamba method over state-of-the-art approaches. We also show that new metrics are more robust for nuclei panoptic segmentation tasks.

## II. RELATED WORK

### A. Nuclei Panoptic Segmentation

In the existing literature, many research papers are identified with semantic or instance segmentation for nuclei images, although some of them include the terms "panoptic segmentation" or "panoptic quality metrics"; however, rarely with panoptic segmentation, which has been summarized in the Introduction section. The major reason for the phenomenon is that there are few pathology image datasets with fully annotated labels that conform to the definition of panoptic segmentation [18]. Hence, due to not being supported by datasets, most of the models in the previous work do not jointly perform semantic and instance segmentation when the panoptic annotation in a pathology image dataset contains image-level tags for "stuff" classes (e.g, uncountable regions like tissue background or necrosis) and the bounding box annotations for "thing" classes (e.g., countable objects like cells/nuclei). Therefore, the panoptic segmentation model that strictly follows the aforementioned definition should simultaneously provide not only tissue (semantic) segmentation but also cell/nuclei (instance) segmentation, or both semantic and instance segmentation for cell/nuclei only when tissue is not the region of interest.

Cell R-CNN [9] employs the backbone ResNet, which integrates semantic and instance branches. The semantic segmentation branch produces segmentation results by classifying each pixel according to its object category. Meanwhile, the feature map branch organizes and provides shared features used by the region proposal network and the instance segmentation components. The instance branch is composed of two parts: one that handles cell positioning and discrimination, and another that generates cell masks. Cell R-CNN is evaluated using F1 score, cell-object Dice, and Hausdorff distance.

As an adaptation of Cell R-CNN, Liu et al. [10] introduce Mask R-CNN for the instance branch to improve the potential feature representation ability for more accurate localization of the cell nuclei through the local and global information fusion. Meanwhile, the semantic feature map is generated from the instance feature masks via the semantic feature fusion module, allowing the final semantic segmentation maps to contain both global and local features. By integrating the semantic branch with the instance branch, the proposed model achieved competitive performance on the evaluation metrics, comparable to Cell R-CNN.

Distinct from the aforementioned unified panoptic segmentation models, there exist a few well-known models in histological image analysis that feature a network architecture capable of simultaneously segmenting and classifying nuclei, i.e., nuclei instance segmentation, such as HoVer-Net [19], HoVer-NeXt [20], and CellViT [21].

### B. Visual Mamba Encoder

Mamba [16] and Mamba-2 [17] were initially introduced to natural language processing for one-dimensional sequences, and then to computer vision for 2D or 3D multidimensional representations. The visual Mamba models for 2D images differ in their design: some feature plain encoders, while others use hierarchical encoders. To give a better visual counterpart to the Mamba block in the encoder, Vision Mamba (ViM) [22] optimizes scan directions for Mamba using the bidirectional Mamba block, while VMamba [13] introduces the visual state space block that employs 2D-slective-scan modules, allowing each pixel in the image to integrate information from all other pixels across different directions. SparX-Mamba [23] is built on VMamba and facilitates a sparse cross-layer connection mechanism within each stage of the visual state space block. MSVMamba [14] improves VMamba via multiscale 2D-slective-scan, and its performance outperforms the above visual Mamba models on image classification, object detection, instance segmentation, and semantic segmentation tasks. Nevertheless, Visual State Space Duality (VSSD) [24] adapts state space duality to a non-causal mode via causal convolution, thereby better handling the intrinsic structural information of image data and achieving slightly better performance than MSVMamba in the semantic segmentation task.

## III. METHODS

Fig. 1 illustrates an overview of the proposed PanopMamba, which incorporates an SSM-based feature-enhanced fusion network with a Mamba-based encoder and a Transformer-based decoder. PanopMamba initially involves the usage of an MSVMamba encoder network to extract multiscale features. These features undergo a feature pyramid and enhancement process to extract deeper features effectively via SSM-based feature-enhanced fusion. Finally, per-pixel and per-segment embeddings are processed by pixel and Transformer decoders separately and concurrently, which also provides a multiscale strategy for successive feature maps and aligns with the encoder and fusion networks.

### A. Pretrained Multiscale VMamba Encoder

MSVMamba has shown a better trade-off between performance and efficiency for many image classification and downstream tasks. The proposed PanopMamba first utilizes an MSVMamba with pre-trained weights to extract multiscale feature representations to improve small object detection in histopathology images. Built on the VMamba, MSVMamba incorporates a multiscale hierarchy module to extract features at different scales or resolutions, capturing both fine-grained details and broader context information.

The input image is first partitioned into patches by a stem module. To produce hierarchical representations, several Multi-Scale State Space (MS3) blocks, interleaved with downsampling modules, are applied to the partitioned patch tokens. In the first stage, an MS3 block is used to produce features with a size of $\frac{H}{4} \times \frac{W}{4}$ tokens. In the second to fourth stages, the number of tokens produced by subsequent MS3 blocks is reduced to $\frac{H}{8} \times \frac{W}{8}$, $\frac{H}{1} \times \frac{W}{1}$, and $\frac{H}{3} \times \frac{W}{3}$ respectively, as the network becomes deeper along with the downsampling modules. The feature maps of F3, F4, and F5 after the 8×, 16×, and 32× downsampling correspond to representations at different scales, capturing low to high-level semantic information. These features are then used as input to the SSM-based feature-enhanced fusion network (in Section III-B).

The MS3 block contains convolutional and Mamba-based modules. It first normalizes the input feature map through LayerNorm (LN) to accelerate the convergence of network training and solve the gradient problem. This is followed by a linear transformation to further extract features. It then utilises Depthwise Convolution (DWConv) [25] layers to reduce computational requirements and enhance feature learning. Next, the MS3 block employs a multiscale 2D scanning (MS2D) module to reduce the total sequence length across four scans. Subsequently, a Squeeze-and-Excitation (SE) [26] block is used for enhanced channel-wise attention by adaptively adjusting the weight of feature channels and thus emphasizing more important channels and suppressing less relevant ones. This can help produce more discriminative features to distinguish overlapping nuclei and improve contrast between different cell types. Further linear transformation enables feature conversion and fusion. To improve channel mixing, the convolutional feedforward network first performs layer normalization to ensure stable data distribution, and 1×1 convolution is applied to adjust the channel.

We apply MS2D on the original and downsampled feature maps. It is an extended Mamba-based module featuring a selective scan mechanism that enables the learning of long-range dependencies while reducing computational cost. MS2D first converts a 2D feature map into a 1D sequence and then scans it with DWConv layers of distinct kernel sizes and strides to enhance its ability to capture global information. The interpolate module plays a crucial role in mitigating the decay in downsampled features. Thus, the MSVMamba encoder with MS2D not only learns fine-grained features in histopathology images but also reduces the number of model parameters compared to other vision SSMs.

We utilise the MSVMamba encoder, pretrained on a large-scale natural image dataset, to leverage its ability to capture generic low- and mid-level features, particularly boundary details crucial for segmentation. Instead of training from scratch, using a pretrained encoder provides ready-to-use feature representations, significantly reducing the required training epochs. Robust feature representations learned on a large dataset allow our model to generalize well even when trained on limited labeled datasets, such as nuclei annotations in histopathology images.

### B. SSM-Based Feature-Enhanced Fusion

To precisely identify and classify nuclear boundaries in regions of interest, we design an SSM-based feature-enhanced fusion network that fuses multiscale feature maps extracted from the MSVMamba encoder. By combining semantic information across multiple scales of the image, this network enables effective handling of objects of varying sizes and shapes, thereby achieving better panoptic segmentation than a pure encoder-decoder architecture.

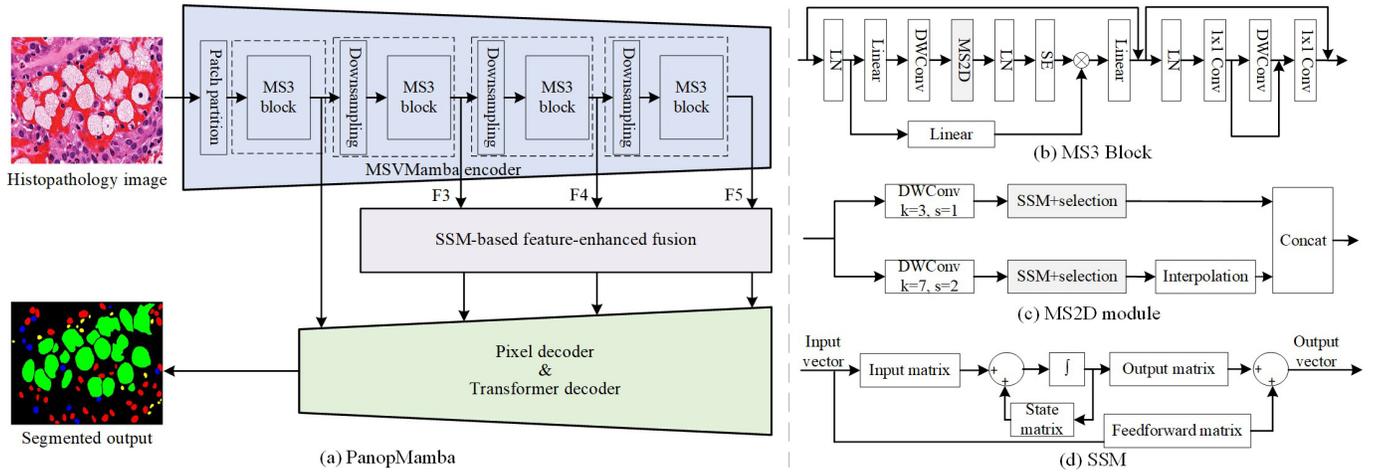

Fig. 1. (a) Overview of PanopMamba. It consists of an MSVMamba encoder, an SSM-based feature-enhanced fusion network, and pixel and Transformer decoders. The MSVMamba encoder features a hierarchical architecture comprising MS3 blocks and downsampling modules. The encoder produces feature maps of hierarchical representations with different resolutions denoted as F3, F4, and F5; For architectural details of SSM-based feature-enhanced fusion network, see Fig. 2; For details of pixel and Transformer decoders, see [6]. (b) Structure of the MS3 block used in the MSVMamba encoder. It consists of LayerNorm (LN), MS2D strategy, an SE block [26], DWConv [25] layers, Convolutional layers (Conv), and Linear transformation (Linear). (c) Structure of MS2D module in MS3 block. It contains an SSM+Selection mechanism from the original Mamba block [13] and a DWConv with parameters of kernel size (k) and stride (s). (d) Structure of SSM used in the MS2D module (highlighted in grey color). A sample histopathology image from MoNuSAC2020 [31] and its segmented output by PanopMamba are shown.

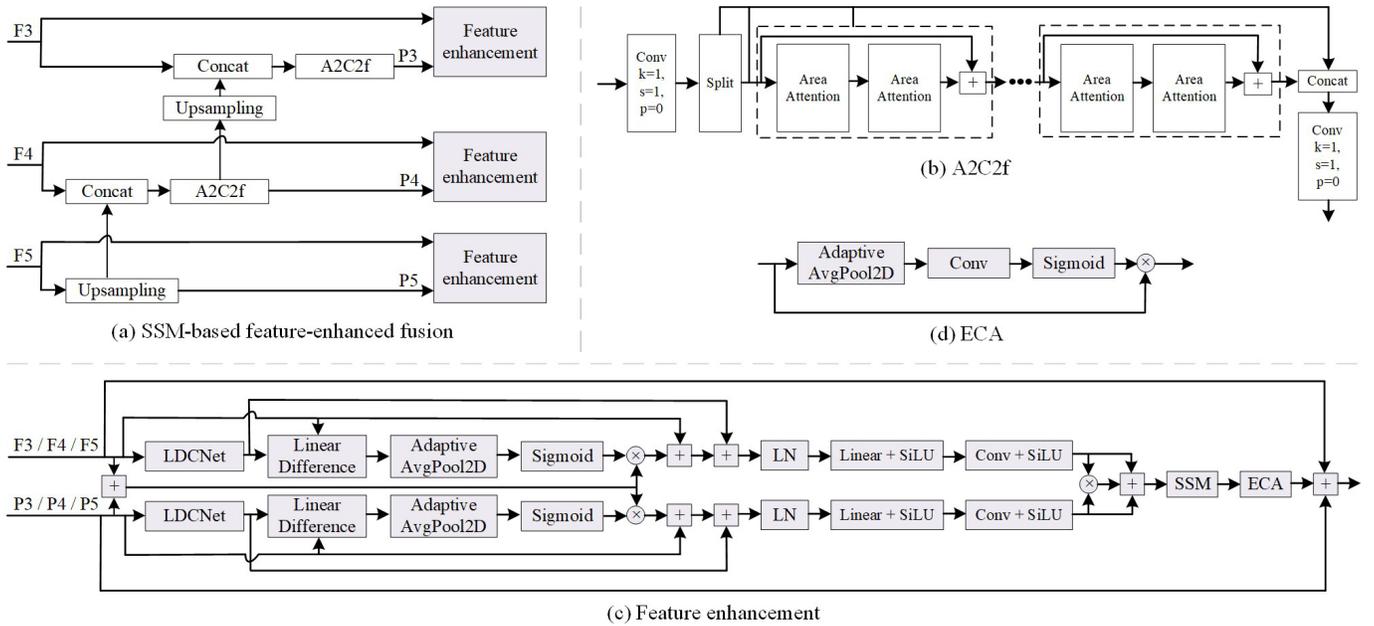

Fig. 2. (a) Detailed architecture of SSM-based feature-enhanced fusion. It features a feature pyramid network structure, incorporating feature enhancement networks (highlighted in purple). The feature maps after the feature pyramid network are denoted as P3, P4, and P5. (b) A2C2f module adopted from YOLOv12 [27] with specific values of kernel size (k), stride (s), and padding (p). (c) The structure of feature enhancement is composed of LDCNet [29], SSM in Mamba block, ECA [30] module, LayerNorm (LN), Sigmoid Linear Unit (SiLU) activation functions, etc. (d) Detail of ECA module in feature enhancement.

Fig. 2 shows the detailed structure of the proposed feature-enhanced fusion network. It assumes the form of a feature pyramid network that generates a hierarchy of feature maps at different levels, namely P3, P4, and P5. The feature pyramid structure consists of convolutional layers with smaller $3 \times 3$ kernels with the bottleneck residual connections to enable the fusion of features from different regions of interest at varying scales (see (b) Area-Attention Cross-stage partial bottleneck with 2 convolutions faster implementation (A2C2f) [27] and its caption in Fig. 2). The area attention module retains the strengths of the multi-head attention mechanism while drastically reducing computational complexity by reducing the spatial dimensions of the key, query, and value. A2C2f integrates attentioned features from multiple stages, enhancing segmentation accuracy, especially for small nuclei, and concatenates all intermediate outputs, which preserves richer feature information and improves learning. High-level input features (e.g., F5) are progressively upsampled to align with them, and then concatenated with the lower-level feature maps (e.g., F4) to ensure that each output feature map contains both fine-grained details and global semantic information. This multiscale feature fusion allows the model to recognize even small objects by effectively combining information from various resolutions.

Furthermore, we introduce a feature enhancement network at each level to refine the features by explicitly modeling dependencies between different scales. This network takes the pyramid feature maps from different scale levels (P3, P4, and P5) and the corresponding input features (F3, F4, and F5) as inputs. The feature enhancement network simultaneously processes feature and pyramid data via two branches that combine CNN-based local context attention with long-range dependency attention over the state space. Inspired by Fusion-Mamba [28], we employ a Learnable Descriptive Convolution Network (LDCNet) [29] at the initial stage of the network to adaptively learn intrinsic textural features from low-level feature maps and suppress irrelevant and redundant features via a difference pooling operation. The Adaptive Average Pooling (Adaptive AvgPool2D) layer is then applied to resize the input feature map to a fixed size while preserving the main features. This is followed by the Sigmoid activation function, which maps the input value to the range [0, 1] and produces a weight coefficient. Internal covariate shift is reduced via the add operations. We introduce the SSM module in the feature enhancement network to achieve a balance between global receptive fields and computational efficiency. SSMs can effectively capture global patterns and long-term correlations in multiscale feature patches. This addresses the limitations of pure CNNs, which have local receptive fields, when extracting features. At the end of this network, the Efficient Channel Attention (ECA) [30] layer facilitates channel-wise attention, which can effectively capture the correlation between channels and learn to select different channel features adaptively, thus optimizing feature fusion.

### C. Pixel and Transformer Decoders

The output feature maps from the SSM-based feature-enhanced fusion network are then decoded to produce segmentation maps. We adopt the pixel and Transformer decoders of Mask2former [5]. The Pixel decoder, as a multiscale decoder, progressively upsamples multiscale features from the output of the SSM-based feature-enhanced fusion network. To generate the final segmentation output, the pixel decoder's mask is combined with its final feature maps to extract foreground features. These foreground maps are then further refined by multiplying them by class information. The Transformer decoder with masked attention generates predictions by decoding pyramid features alongside masked query features. The combination of pixel and Transformer decoders plays a crucial role in producing fine-grained, high-resolution feature maps (per-pixel embeddings) that preserve an image's detailed spatial information.

Due to the incorporation of vision state space modeling in both the feature extraction of MSVMamba Encoder and feature fusion of SSM-based feature-enhanced fusion network, our model outperforms Mask2former in terms of segmentation of small and overlapping objects. Without the strong multiscale encoding and enhanced fusion, the decoder cannot generate segmentation maps that capture small-scale information from compressed encodings.

## IV. EXPERIMENTAL RESULTS

### A. Datasets

To evaluate the proposed PanopMamba model, we used two benchmark H&E stained histological image datasets, where images use H&E dyes that make it easier to see different parts of the cell under a microscope as a common laboratory method: (i) Multi-Organ Nuclei Segmentation and Classification Challenge (MoNuSAC2020) [31] composed of 4 types of cells from 4 different human organs and (ii) A fully-annotated dataset for Nuclei Instance Segmentation (NuInsSeg) [32] from 31 human and mouse organs. In particular, the two datasets contain only various nuclei instances and semantic segmentation, but do not include tissue panoptic segmentation; therefore, the tissue in the task is identified as background. The cell types in MoNuSAC2020 include epithelial cells, lymphocytes, macrophages, and neutrophils from the lung, kidney, breast, and prostate, while NuInsSeg comprises 31 categories. We randomly split them into train and test sets at an 8:2 ratio and converted their annotations to the MS COCO panoptic annotation format, respectively.

Besides, our color generation for panoptic data visualization is different from COCO's because the typical COCO panoptic annotations are stored as polygons with decimal code representing their vertices for different classes of instances. Due to the incompatibility of standard COCO/ADE20K/Cityscapes panoptic annotations and unknown additional requirements for input annotation formats, some panoptic segmentation models in the existing literature cannot be run with our converted typical formats.

### B. Implementation Details

Our proposed model is built on the Python package MM-Segmentation v1.2.2 (released on 14 Dec 2023), with the pretrained MSVMamba-S segmentation checkpoint serving as the backbone. We adopted MSVMamba-Tiny with pretrained checkpoints on the ImageNet-1K dataset; for details of pre-training settings, see [14]. The proposed model is conducted using PyTorch with an NVIDIA® GeForce RTX® 4090 GPU with 24 GB of memory. For a fair comparison, all the proposed and competitive models were run for 180,000 iterations in the training stage. AdamW optimizer is adopted to optimize our proposed model, the learning rate is set as 0.0001, and weight decay is set to 0.05.

PERFORMANCE COMPARISON OF STATE-OF-THE-ART MODELS AND OUR PROPOSED PANOPMAMBA ON MoNuSAC2020

| Model | PQ | $m$PQ+ | $b$PQ | $i$PQ | $w$PQ | $fw$PQ | $R^2$ |
|---|---|---|---|---|---|---|---|
| MaskFormer [12] | 36.37 | 36.37 | 34.97 | 28.44 | 39.25 | 34.05 | 19.69 |
| Panoptic SegFormer [4] | 35.82 | 35.82 | 34.09 | 27.92 | 38.51 | 33.87 | 18.97 |
| Mask2Former [5] | 39.76 | 39.76 | 38.60 | 33.05 | 42.19 | 37.99 | 57.50 |
| OneFormer [7] | 31.35 | 31.35 | 29.85 | 17.57 | 32.75 | 19.95 | 27.66 |
| MP-Former [6] | 31.98 | 31.98 | 31.13 | 12.93 | 31.79 | 16.22 | 13.41 |
| HoVer-Net [19] | 31.15 | 31.15 | 30.43 | 22.98 | 33.42 | 27.27 | 35.18 |
| HoVer-NeXt [20] | 51.33 | 51.33 | 51.09 | 44.21 | 57.15 | 48.98 | 65.59 |
| CellViT [21] | 27.08 | 27.08 | 26.97 | 25.63 | 27.70 | 28.53 | 26.52 |
| **PanopMamba (Ours)** | **73.11** | **73.11** | **72.14** | **74.28** | **75.38** | **74.56** | **86.93** |

All the results represent the average values (%) of the metrics across all classes. The best results are shown in bold.

PERFORMANCE COMPARISON OF STATE-OF-THE-ART MODELS AND OUR PROPOSED PANOPMAMBA ON NuInsSeg

| Model | PQ | $m$PQ+ | $b$PQ | $i$PQ | $w$PQ | $fw$PQ | $R^2$ |
|---|---|---|---|---|---|---|---|
| MaskFormer | 49.82 | 46.70 | 48.86 | 52.38 | 56.07 | 53.43 | 65.19 |
| Panoptic SegFormer | 17.96 | 16.84 | 17.68 | 9.54 | 16.97 | 9.63 | 7.18 |
| Mask2Former | 47.35 | 44.39 | 46.47 | 50.02 | 52.23 | 50.97 | 75.55 |
| OneFormer | 16.91 | 15.86 | 16.37 | 13.99 | 16.71 | 14.22 | 16.37 |
| MP-Former | 44.73 | 41.94 | 44.12 | 42.63 | 48.15 | 41.52 | 43.96 |
| HoVer-Net | 24.89 | 23.33 | 24.48 | 20.58 | 26.58 | 21.41 | 30.63 |
| HoVer-NeXt | 29.68 | 25.97 | 29.17 | 29.08 | 31.92 | 29.48 | 21.39 |
| CellViT | 19.94 | 18.69 | 19.42 | 10.05 | 20.42 | 13.09 | 39.02 |
| **PanopMamba (Ours)** | **73.68** | **69.08** | **73.01** | **79.51** | **80.69** | **79.96** | **97.05** |

All the results represent the average values (%) of the metrics across all classes. The best results are shown in bold.

## C. Evaluation Metrics

Foucart et al. [1], [2] have pointed out the robustness issue of the Panoptic Quality (PQ) metric proposed by [18], especially for small objects such as cell nuclei, due to IoU's sensitiveness, whereas the use of the IoU as a matching rule and as a segmentation quality measure within PQ. However, van Heusden and Marx [33] show that IoU with altered conditions can be used in determining the correctness of a prediction against a gold standard. In the existing literature, boundary PQ ($b$PQ) [34] and multiclass PQ ($m$PQ+) [35] have been proposed as solutions to measure segmentation evaluation with focus on previously overlooked characteristics, such as boundary errors and unmatched instances, which are more sensitive than the standard evaluation protocols. The value of $m$PQ+ equals the mean vanilla PQ if there are no unmatched ground truth and predicted instances, when ground truth and predicted instances of each image and category within the dataset are matched into pairs. Additionally, the multiclass coefficient of determination ($R^2$) [35] between the predicted and true nuclei counts is defined as the coefficient of determination for multiclass instance/panoptic segmentation, indicating the proportion of the predicted nuclei counts that are correct for all images and categories.

To further investigate this issue, we propose three new metrics for robustly evaluating panoptic segmentation. Inspired by recent new metrics such as fine-grained image-level IoU at a per-image level ($i$IoU) [36], weighted IoU ($w$IoU) [37], and frequency-weighted IoU ($fw$IoU) [38], [39], we introduce three alternative PQ metrics, $i$PQ, $w$PQ, and $fw$PQ, as complements of PQ to provide a more comprehensive evaluation for panoptic segmentation tasks, which take into account detection of small objects, ambiguous boundaries, or imbalanced classes, respectively. Specifically, PQ averaged on a per-image basis ($i$PQ) yields a per-image-per-class score through averaging IoUs by class and then by image to mitigate class/size imbalance; boundary-weighted PQ ($w$PQ) evaluates both the contour and region of the segments; frequency-weighted PQ ($fw$PQ) weighs the frequency of occurrence of each class. The proposed metrics are defined as:

$$i/w/fwPQ = \frac{i/w/fwIoU}{|TP| + \frac{1}{2}|FP| + \frac{1}{2}|FN|} \quad (1)$$

where $i$, $w$, and $fw$ target different altered conditions for IoU and PQ. $i$PQ is a per-image score that reduces the size imbalance from dataset-level to image-level. In multiclass segmentation of $i$PQ computation, any class absent from the ground truth is assigned a value of null for that image, even if it was detected in the prediction. The boundary-based metrics $b$PQ and $w$PQ both have a boundary weight to set. In our experiments, we set the contour parameter $d = 0.02$ for $b$PQ and the boundary importance factor $a = 10$ for $w$PQ. $fw$PQ is a class-sensitive metric for datasets with highly imbalanced class distributions. True Positive (TP) represents the number of instances of each image and category where the model correctly predicted the positive class. False Positive (FP) or False Negative (FN) represents the number of instances where the model incorrectly predicted the positive or negative class. The values above and below the fraction represent segmentation and recognition/detection quality, which approach 1.

## D. Quantitative Results

### 1) Overall Average Results:
We perform extensive comparison experiments with the state-of-the-art panoptic segmentation methods to demonstrate the validity of our proposed

TABLE III

Per-class results across various PQ metrics of Mask2Former and our proposed PanopMamba on MoNuSAC2020

| Model | Category | PQ | $mPQ+$ | $bPQ$ | $iPQ$ | $uPQ$ | $fwPQ$ |
|---|---|---|---|---|---|---|---|
| Mask2Former | Epithelial | 53.30 | 53.30 | 53.30 | 40.84 | 57.69 | 54.08 |
| | Lymphocyte | 50.76 | 50.76 | 50.74 | 48.98 | 54.03 | 52.75 |
| | Neutrophil | 40.30 | 40.30 | 37.77 | 33.62 | 44.71 | 35.14 |
| | Macrophage | 14.67 | 14.67 | 12.58 | 8.78 | 12.33 | 9.98 |
| PanopMamba | Epithelial | **78.46** | **78.46** | **78.42** | **81.35** | **81.48** | **82.62** |
| | Lymphocyte | **79.26** | **79.26** | **79.21** | **79.82** | **80.89** | **81.47** |
| | Neutrophil | **76.42** | **76.42** | **73.70** | **73.91** | **78.48** | **73.06** |
| | Macrophage | **58.28** | **58.28** | **57.20** | **62.05** | **61.80** | **61.11** |

The unit of values are percentage (%). The best results are shown in bold.

TABLE IV

Per-class results in vanilla PQ and its component metrics of Mask2Former and our proposed PanopMamba on MoNuSAC2020

| Model | Category | PQ | SQ | RQ |
|---|---|---|---|---|
| Mask2Former | Epithelial | 53.31 | 78.70 | 67.74 |
| | Lymphocyte | 50.76 | 77.03 | 65.90 |
| | Neutrophil | 40.30 | 79.46 | 50.72 |
| | Macrophage | 14.67 | 77.52 | 18.92 |
| **PanopMamba** | Epithelial | **78.46** | **89.06** | **88.10** |
| | Lymphocyte | **79.26** | **90.72** | **87.37** |
| | Neutrophil | **76.42** | **91.56** | **83.47** |
| | Macrophage | **58.27** | **81.59** | **71.43** |

The unit of values are percentage (%). The best results are shown in bold.

PanopMamba. The highest vanilla mean PQ (i.e., vanilla mPQ, which is the average value of vanilla PQ values for each image; hereinafter called PQ) on the MoNuSAC2020 challenge leaderboard is 0.6119 (i.e., 61.19%), while the NuInsSeg segmentation benchmark result yields a PQ of 0.513 (i.e., 51.3%). Because the algorithms that participated in the challenge were unavailable, they could not be included in our analysis. We select state-of-the-art models in panoptic segmentation to validate on two datasets, including MaskFormer [12] with ResNet50 backbone, Panoptic SegFormer [4] with ResNet50, Mask2former [5] with Swin-S, OneFormer [7] with Swin-L, and MP-Former [6] with Swin-L, as well as nuclei instance segmentation models HoVer-Net [19], HoVer-NeXt [20], and CellViT [21]. The experimental results are based on robust evaluation metrics, including vanilla PQ and its variants $mPQ+$, $bPQ$, $iPQ$, $uPQ$, and $fwPQ$, as well as $R^2$, and our proposed variants are consistent with their semantic segmentation metric IoUs.

The quantitative results are presented in Table I for the MoNuSAC2020 dataset. Clearly, the proposed model outperforms all others across various panoptic segmentation evaluation metrics. Our proposed PanopMamba achieves 73.11% mean PQ values and 86.93% predicted nuclei counts on MoNuSAC2020, representing an improvement of approximately 33.35% compared to the second-best model, Mask2Former. In terms of vanilla PQ variants, PanopMamba with a feature fusion network outperforms existing encoder-decoder models, achieving the best benchmark results on the MoNuSAC2020 dataset across alternative PQ metrics.

Table II presents the average results for all PQ metrics on NuInsSeg. Our proposed PanopMamba still achieves better scores across all metrics, even though all the models achieve better panoptic segmentation quality on the NuInsSeg dataset than the other one, because there are more medium- and large-sized nuclei in NuInsSeg, which generally make them easier to identify and segment by models. From the results, PanopMamba demonstrates outstanding performance in nuclei panoptic segmentation, achieving a noteworthy PQ of 73.68%, which clearly surpasses state-of-art models, such as Mask-Former, which achieves only 49.82%.

*2) Per-Class Results:* To illustrate the superior performance of PanopMamba, we present per-class results for nuclei of four cell types (epithelial, lymphocyte, neutrophil, and macrophage) across various PQ metrics in Table III. The results for each class are consistent with those of the overall average. We further compare the per-class results of the second-best panoptic segmentation models, Mask2Former and PanopMamba, using PQ submetrics. Given the same results for $mPQ+$ as PQ on MoNuSAC2020, we only select PQ's component metrics, Segmentation Quality (SQ) and Recognition Quality (RQ), for analysis. It is evident that PanopMamba excels in both SQ and RQ components, as shown in Table IV.

*3) Robustness:* The robustness of the proposed Panop-Mamba is validated across various panoptic segmentation quality metrics, while the distinctiveness of various PQ variants is demonstrated by evaluating them on different models. The consistency of PanopMamba's better results compared to state-of-the-art competitors, across all classes and each class, on six distinct metrics, is notable for its robustness in nuclei panoptic segmentation.

On the other hand, our proposed alternative PQ metrics, $iPQ$, $uPQ$, and $fwPQ$, have been successfully designed for evaluating the quality of panoptic segmentation. The PQ variants address concerns about IoUs that involve altered conditions across different settings, where IoU determines the correctness of a prediction against a gold standard. According to the quantitative results for average all-class and per-class nuclei, the alternative PQ metrics can be distinguished by their different definitions. Their values typically follow the ranking: $uPQ \geq fwPQ \geq iPQ \geq PQ \geq mPQ+$ or $bPQ$, which can be distinguished by their unique value ranges. It appears that $bPQ$ is the strict metric for evaluating nuclei panoptic segmentation among the six, as its high value is more challenging to achieve.

### E. Abalation Study

*1) Ablation Experiments on Each Component:* We conducted ablation experiments on the proposed panoptic segmentation method, PanopMamba, on the MoNuSAC2020 dataset by replacing the MSVMamba encoder with a VMamba [13] encoder and removing either the SSM-based fusion network or the pixel decoder. As shown in Table V, integration of either MSVMamba Encoder or our proposed SSM-based feature-enhanced fusion can improve the performance of nuclei panoptic segmentation by a clear margin. The performance decreases significantly by approximately 0.20 without SSM-based fusion, indicating its importance in the proposed method.

The results of the ablation experiment on each component in the proposed PanopMamba demonstrate that feature-enhanced fusion using the encoder-decoder framework consistently outperforms the same approach without it for nuclei panoptic segmentation. Our proposed SSM-based feature-enhanced fusion combines the strengths of both feature pyramid fusion techniques and a feature enhancement mechanism that dynamically perceives differences among multiscale features, further validating the advantage of state space models. In the proposed feature fusion, LDCNet effectively captures fine-grained textural patterns, and SSMs introduce hidden states to remember long-range dependencies, thus making it well-suited for the nuclei segmentation task.

*2) Ablation Experiments on Visual Mamba Encoder:* We dissect the MSVMamba encoder in the proposed model and evaluate the effectiveness of different visual Mamba encoders, comparing them with the performance of using only the unique encoder. All the encoders for comparison, VMamba-T [13], SparX-Mamba-B [23], VSSD-Base [24], and MSVMamba-S [14], are pretrained on a large-scale dataset, and the pretraining setups follow a similar experimental setting described in their respective papers. We select the largest size provided in their code repositories, despite the different naming conventions. We observe that the results of MSVMamba-Tiny in Table VI surpass those of other Mamba-based encoders with the same fusion and decoder components.

Since one of the challenges of nucle segmentation is their small size, multiscale feature extraction is vital in this context. This approach is crucial because small nuclei often lack sufficient feature representation at a single scale, making them difficult to detect. By combining features from various scales, models can better learn discriminative features and improve detection accuracy, especially in complex scenes. MSVMamba replaces the 2D-selective-scan with a multiscale 2D-selective-scan to further introduce a hierarchical design within a single layer and integrates squeeze-excitation attention after the multiscale 2D scanning, alleviating decay along scanning routes in downsampled maps and enhancing the ability to capture global information.

*3) Ablation Experiments on Different Fusion Networks:* We investigate the effectiveness of the proposed SSM-based feature-enhanced fusion and FusionMamba [28], a comprehensive framework comprising feature extraction, feature fusion, and feature reconstruction modules for fusion tasks involving two different image modalities. To adapt to the proposed PanopMamba, we select only the dynamic feature enhancement module (dynamic feature fusion and cross-modal fusion Mamba modules) from the FusionMamba network to replace our proposed fusion network in the proposed model, as the application has a different purpose. According to the compar-isons in Table VII, our proposed fusion network outperforms the other Mamba-based one.

*4) Ablation Experiments on Feature Pyramid Network:* We analyze the importance of the feature pyramid network among cross-stage partial bottleneck with 3 convolutions (C3) in YOLOv5 [40], C3 with kernel value equal 2 (C3k2) in YOLO11 [41], and A2C2f in YOLOv12 [27]. Since YOLOv5, the convolutional block, composed of a 2D convolutional layer, a 2D batch normalization, and a SiLU activation function, has been introduced in all YOLO versions in recent years. C3k2 replaces the bottlenecks with a single C3k, and A2C2f marks the full integration of attention mechanisms into YOLOv12.

The results in Table VIII show that the feature pyramid network in the proposed feature-enhanced fusion using A2C2f has significantly better panoptic qualities than C3 and C3k2.

From the above results, it can be observed that the attention mechanisms can help improve segmentation accuracy in the feature pyramid network. Alongside lightweight area attention with simple reshape operation, A2C2f achieves efficient global and local semantic information while maintaining real-time speed and improving robustness and performance. Combining the area attention mechanisms, the feature pyramid network enhances its feature representation and contextual information for segmenting small nuclei. Moreover, it refines feature fusion within the feature pyramid network, resulting in more robust and accurate nucleus segmentation.

### F. Qualitative Results

As illustrated in Fig. 3, Mask2Former and OneFormer even wrongly classify some of the nuclei as another category compared to the ground truth of the sample image in MoNuSAC2020, which leads to a lower precision for panoptic quality than the proposed PanopMamba. In the sample comparison in NuInsSeg, our proposed method effectively segments small nuclei within cells by leveraging the Mamba-based pretrained encoder and SSM-based feature-enhanced fusion, whereas the competing models, using Mask2former and OneFormer, fail to accurately distinguish these nuclei in shape and boundary from the tissue background. The visualization of segmented samples demonstrates that PanopMamba has advantages in segmenting small nuclei.

## V. Conclusion

We developed a Mamba/SSM-based network, Panop-Mamba, to address the issues of nuclei panoptic segmentation, leveraging long-range information captured and fused by the Mamba-based encoder and fusion network. This model integrates MSVMamba, pretrained on a large-scale natural image dataset, as its feature-extraction component. MSVMamba is an efficient vision Mamba encoder, especially in handling small objects, that adeptly introduces the multiscale technique on Mamba-based 2D scanning using depthwise convolutions with different kernel and stride values. The Mamba-based multi-scale encoder we employ combines the strength of long-range information extraction and computation reduction, addressing the trade-off between model accuracy and computational efficiency.

Ablation experiments considering different combinations of PanopMamba components on MoNuSAC2020

| Method | PQ | mPQ+ | bPQ | iPQ | uPQ | fuPQ | R$^2$ |
|---|---|---|---|---|---|---|---|
| PanopMamba (w/o MSVMamba Encoder) | 54.58 | 54.58 | 52.73 | 57.38 | 61.71 | 58.68 | 76.36 |
| PanopMamba (w/o SSM-based Fusion) | 37.83 | 37.83 | 36.27 | 38.52 | 38.22 | 38.82 | 53.19 |
| PanopMamba (w/o Pixel Decoder) | 59.46 | 59.46 | 58.32 | 60.74 | 61.12 | 61.02 | 67.81 |

All the results represent the average values (%) of the metrics across all classes. w/o denotes without.

TABLE VI

Ablation experiments on different visual Mamba encoders in our proposed PanopMamba on MoNuSAC2020

| Method | PQ | mPQ+ | bPQ | iPQ | uPQ | fuPQ | R$^2$ |
|---|---|---|---|---|---|---|---|
| VMamba-T [13] | 54.58 | 54.58 | 52.73 | 57.38 | 61.71 | 58.68 | 59.36 |
| SparX-Mamba-B [23] | 46.99 | 46.99 | 45.46 | 48.60 | 51.96 | 48.62 | 57.49 |
| VSSD-Base [24] | 62.91 | 62.91 | 61.65 | 65.73 | 68.67 | 67.72 | 85.39 |
| **MSVMamba-S [14] (Ours)** | **73.11** | **73.11** | **72.14** | **74.28** | **75.38** | **74.56** | **86.93** |

All encoders with their largest size are used, as provided in their respective code repositories.
The best results are shown in bold.

TABLE VII

Ablation experiments on different fusion networks in our proposed PanopMamba on MoNuSAC2020

| Method | PQ | mPQ+ | bPQ | iPQ | uPQ | fuPQ | R$^2$ |
|---|---|---|---|---|---|---|---|
| FusionMamba [28] | 64.35 | 64.35 | 63.02 | 65.38 | 66.89 | 65.77 | 74.93 |
| **Proposed SSM-Based Feature-Enhanced Fusion (Ours)** | **73.11** | **73.11** | **72.14** | **74.28** | **75.38** | **74.56** | **86.93** |

All the results represent the average values (%) of the metrics across all classes. The best results are shown in bold.

TABLE VIII

Ablation experiments on different feature pyramid modules in our proposed SSM-based feature-enhanced fusion on MoNuSAC2020

| Method | PQ | mPQ+ | bPQ | iPQ | uPQ | fuPQ | R$^2$ |
|---|---|---|---|---|---|---|---|
| C3 [40] | 62.39 | 62.39 | 60.86 | 65.18 | 69.33 | 67.04 | 84.74 |
| C3k2 [41] | 57.86 | 57.86 | 55.71 | 60.67 | 65.65 | 61.83 | 75.86 |
| **A2C2f [27] (Ours)** | **73.11** | **73.11** | **72.14** | **74.28** | **75.38** | **74.56** | **86.93** |

All the results represent the average values (%) of the metrics across all classes. The best results are shown in bold.

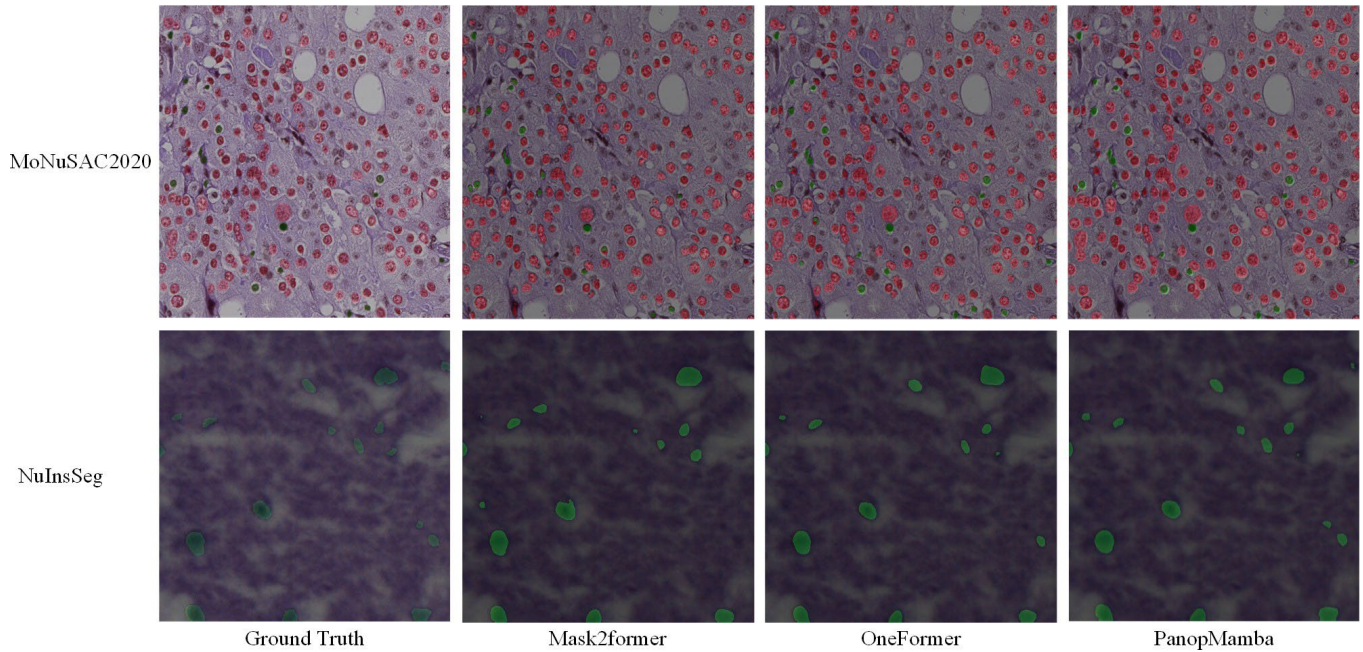

Fig. 3. Qualitative comparisons of Mask2former, OneFormer, the proposed PanopMamba, and the Ground Truth on the same sample in the test sets. The image sample from MoNuSAC2020 in the upper row is TCGA-YL-A9WY-01Z-00-DX1-1.jpg. The lower row is the file human_brain_10.jpg from NuInsSeg.

More importantly, we propose an SSM-based feature-enhanced fusion network that integrates a feature pyramid network with feature enhancement to capture spatial and long-term dependencies, thereby significantly improving the performance of nuclei panoptic segmentation. The results of the ablation study demonstrate that the proposed fusion network achieves at least a 20% increase in panoptic quality. The integration of feature enhancement with an SSM, followed by optimization through an attention module, allows for a robust rebalancing of global and local features. PanopMamba ensures comprehensive global and local receptive field coverage while minimizing computational load. The ablation experiment on components of the proposed method indicates that incorporating feature-enhanced fusion into the encoder-decoder consistently outperforms popular models of existing architectures.